\documentclass[12pt,preprint]{aastex}

\shorttitle{EVOLUTION OF DIFFUSE RADIO SOURCES}

\shortauthors{KUO, HWANG, \& IP}

\begin{document}

\title{The Evolution of Diffuse Radio Sources in Galaxy Clusters}

\author{Ping-Hung Kuo, Chorng-Yuan Hwang, and Wing-Huen Ip}

\affil{Institute of Astronomy, National Central University, Chung-Li 32054,
Taiwan}
\email{d882001@astro.ncu.edu.tw; hwangcy@astro.ncu.edu.tw;
wingip@astro.ncu.edu.tw}

\begin{abstract}

We investigate the evolution and number distribution of radio halos in galaxy
clusters. Without re-acceleration or regeneration, the relativistic electrons
responsible for the diffuse radio emission will lose their energy via
inverse-Compton and synchrotron losses in a rather short time, and radio halos
will have lifetimes $\sim$ 0.1 Gyr. Radio halos could last for $\sim$ Gyr if a
significant level of re-acceleration is involved. The lifetimes of radio halos
would be comparable with the cosmological time if the radio-emitting electrons
are mainly the secondary electrons generated by pion decay following
proton-proton collisions between cosmic-ray protons and the thermal
intra-cluster medium within the galaxy clusters. Adopting both observational
and theoretical constraints for the formation of radio halos, we calculate the
formation rates and the comoving number density of radio halos in the
hierarchical clustering scheme. Comparing with observations, we find that the
lifetimes of radio halos are $\sim$ Gyr. Our results indicate that a
significant level of re-acceleration is necessary for the observed radio halos
and the secondary electrons may not be a dominant origin for radio halos.

\end{abstract}


\section{INTRODUCTION}\label{secint}

Diffuse radio emission from galaxy clusters is a rare phenomenon. These radio
sources, which usually possess large sizes and steep spectra, are called radio
halos if they permeate the cluster centers and radio relics if they are located
in cluster peripheral regions. Observations found that radio halos only exist
in the clusters that show X-ray substructures \citep{fer00}. Since a galaxy
cluster possessing X-ray substructures indicates that it is under ongoing
merging, it is expected that the formation of radio halos is closely related to
the merging process of galaxy clusters.

The diffuse radio emission of galaxy clusters is believed to be produced by the
synchrotron radiation of relativistic electrons. Nonetheless, the sources of
these relativistic electrons are still unclear. Cluster merging is a very
violent event and releases a large amount of energy ($\sim10^{64}$ ergs); this
leads cluster mergers to be a very favorable mechanism for the production of
the relativistic particles. However, relativistic electrons lose energy on the
time scale of order $\sim10^{8}$ years because of inverse Compton and
synchrotron losses; this suggests that without re-acceleration radio halos in
galaxy clusters might be transient features associated with a major merger and
would have lifetimes $\sim$ 0.1 Gyr \citep{dey92,tri93}.

The numerical simulations of cluster mergers \citep[e.g.,][]{min00,ric01}
showed that the intra-cluster medium (ICM) is seriously disturbed by merging.
Violent turbulence generated by mergers must play an important role in the
re-acceleration of relativistic electrons \citep{sar01}. Considering
re-acceleration for the relativistic electrons, a two-phase model proposed by
\citet{bru01} successfully reproduces the radial steepening of the spectral
index, the radio spectrum steepening at high frequencies, and the HXR excess in
the Coma cluster \citep*[see also][]{sch87,lia02}. On a similar study,
\citet*{kuo03} showed that the ``age" of Coma C might be $\sim$ 1 Gyr; this
indicates that the lifetimes of radio halos could be $\sim$ Gyr if a
significant level of re-acceleration is involved.

The secondary electron model first proposed by \citet{den80}
assumes that relativistic electrons are produced from the pion
decay following collisions between the cosmic-ray protons and the
thermal ions of the ICM. It has been recognized that the diffusion
time of cosmic-ray protons is comparable with the cosmological
time, so cosmic-ray protons are confined within galaxy clusters
for the lifetimes of the clusters \citep*{vol96,ber97,sch87}. If
radio halos are formed from the secondary electrons
\citep[e.g.,][]{bla99,min01a}, their lifetimes would be comparable
with the cosmological time. The significantly different time
scales of radio halos could have discernible effects on their
number distribution and thus could discriminate on the origins of
radio halos.

The formation rates of radio halos in galaxy clusters can be estimated from the
hierarchical model. \citet[hereafter PS]{pre74} derived a mass function to
evaluate the comoving number density of bound virialized objects, but this
function does not specify the formation epochs of the objects. To solve this
problem, \citet[hereafter LC]{lac93} derived a distribution function of
formation epochs by using the merger probabilities in the framework of PS
formalism \citep[see also][]{bon91,bow91}. Based on the formalism of LC,
\citet[hereafter KS]{kit96a} proposed another distribution function of
formation epochs in a similar but slightly different manner. We follow the
formalism of KS but modify it to suit the situation for forming radio halos.

In this paper, we investigate the evolution and number distribution of radio
halos with different lifetime scales. Three typical lifetimes that are
representatives of different origins for radio halos are considered (i.e. 0.1
Gyr, 1 Gyr, and the cosmological time). We estimate the different number
distributions and compare the results with observations to determine the valid
models that are responsible for the origin of radio halos.

This paper is planned as follows. In $\S$~\ref{secfc}, we discuss the
conditions for radio halos forming in galaxy clusters. In $\S$~\ref{secfor}, we
describe the methods based on the formalism proposed by KS for calculating the
formation rates and comoving number density of radio halos. In
$\S$~\ref{seccom}, we compare the modelling results of radio halos with
different lifetimes with observations, and in $\S$~\ref{seccon} we give our
discussion and conclusions.

\section{FORMATION CRITERIA}\label{secfc}

Radio halos are not found in low X-ray luminosity clusters and only present in
massive clusters with high X-ray luminosity and high temperature
\citep*{gio99}. This fact indicates that there might be a threshold mass for
galaxy clusters to form radio halos. It has been recognized that radio halos
are strongly correlated with cluster mergers \citep{fer00}. Nonetheless, many
cluster mergers do not possess radio halos. \citet{buo01} studied the dynamical
states of clusters possessing radio halos and found that radio halos form
preferentially in massive clusters experiencing violent mergers that have
seriously disrupted the cluster core. Disrupting the cluster core in the
merging process should be an important factor for forming radio halos.
According to above discussion, we assume two criteria for a cluster merger to
form a radio halo: (1) the cluster mass must be greater than or equal to a
threshold mass, and (2) the merging process must be violent enough to disrupt
the cluster core. Under these two conditions, mergers might generate sufficient
primary electrons or secondary electrons from cosmic-ray protons to form radio
halos.

We use observational data \citep{gio99,fer00,kem01} to determine the threshold
mass. We found that the A548b cluster has the lowest temperature $\sim$ 2.4 keV
\citep{gio99}, which is presumed to have the smallest mass from the well-known
mass-temperature relation for galaxy clusters \citep*[e.g.,][]{mul00,ros02}.
Using the observed mass-temperature relationship \citep*{evr96,hor99}
\begin{equation}\label{Emt}
  M= 5 \times 10^{13} h^{-1}
  \left(\frac{T_{\mathrm{X}}}{\mathrm{1 keV}}\right)^{1.5}
  M_{\odot},
\end{equation}
we found that the A548b cluster has a mass of $\approx 2.7 \times 10^{14}
M_{\odot}$ for $h=0.7$ or $3.7 \times 10^{14} M_{\odot}$ for $h=0.5$. Since
very few clusters with lower temperatures have been found to possess radio
halos, the threshold mass may be on the order of $\sim 10^{14} M_{\odot}$. We
choose $10^{14} M_{\odot}$ to be the threshold mass $M_{\mathrm{th}}$ for
cluster mergers to form radio halos. For comparison, we also consider different
$M_{\mathrm{th}}$ ($5 \times 10^{13} M_{\odot}$ and $5 \times 10^{14}
M_{\odot}$) in our calculation.

The condition for the disruption of cluster cores have been determined from
dynamical simulation. \citet*{sal98} found that a cluster with mass $M$
experiencing a merging process would disrupt its core structure when the
relative mass increase $\triangle\!M / M$ exceeds a certain threshold
$\triangle_{\mathrm{m}} \equiv (\triangle\!M / M)_{\mathrm{threshold}} = 0.6$.
We use these two conditions, $M_{\mathrm{th}}=10^{14} M_{\odot}$ and
$\triangle_{\mathrm{m}}=0.6$, as the criteria for cluster mergers to form radio
halos.

\section{FORMULATION}\label{secfor}

\subsection{Formation Rates}\label{secfr}

The Press-Schechter mass function (PS) is used to model the cluster number
density and its evolution. The comoving number density of clusters in the mass
range $M \sim M+dM$ at time $t$ is given by
\begin{equation}\label{Eps}
  n_{\mathrm{ps}}(M,t)dM=\sqrt{\frac{2}{\pi}} \frac{\rho_{0}}{M}
  \frac{\delta_{c}(t)}{\sigma^{2}(M)} \left|\frac{d\sigma(M)}{dM}\right|
  exp\left[-\frac{\delta^{2}_{c}(t)}{2\sigma^{2}(M)}\right] dM,
\end{equation}
where $\rho_{0}$ is the present mean density of the universe, $\delta_{c}(t)$
is the critical density threshold for a spherical perturbation to collapse by
the time $t$, and $\sigma(M)$ is the present rms density fluctuation smoothed
over a region of mass $M$. We adopt the expressions of $\delta_{c}(t)$
summarized in \citet{ran01} for different cosmological models. For $\sigma(M)$,
we use an approximate formula proposed by \citet{kit96b} for the cold dark
model fluctuation spectrum and choose the value of $\sigma_{8}$ from the
$\Omega_{0}-\sigma_{8}$ constraint derived from the present cluster abundance:
$\sigma_{8}\Omega_{0}^{0.45}=0.53$ (for $\Omega_{\Lambda}=0$) and
$\sigma_{8}\Omega_{0}^{0.53}=0.53$ (for $\Omega_{0}+\Omega_{\Lambda}=1$)
\citep{pen98}. The parameter $\Omega_{0} \equiv \rho_{0}/\rho_{c}$ is the ratio
of the present mean density to the critical density $\rho_{c}=3H_{0}^{2}/(8 \pi
G)$ and $\Omega_{\Lambda} \equiv \Lambda/(3H_{0}^{2})$, where $\Lambda$ is the
cosmological constant. The Hubble constant is defined as $H_{0}=$ 100$h$ km
s$^{-1}$ Mpc$^{-1}$. Three different cold dark matter (CDM) models are
considered in our analysis: a standard model (SCDM) ($\Omega_{0}=1,
\Omega_{\Lambda}=0, h=0.5, \Gamma=0.5$), an open model (OCDM) ($\Omega_{0}=0.3,
\Omega_{\Lambda}=0, h=0.7, \Gamma=0.2$), and a low-density flat model
($\Lambda$CDM) ($\Omega_{0}=0.3, \Omega_{\Lambda}=0.7, h=0.7, \Gamma=0.2$),
where $\Gamma$ is the shape parameter defined by \citet{sug95}.

The formation rate for an object with mass $M_{f}$ ($M_{f} > M_{i}$) formed
from initial mass between $M_{i}$ and $M_{i}+dM_{i}$ in unit time at $t$ is
given by (LC; KS)
\begin{equation}\label{Emr}
  r(M_{i}\rightarrow M_{f};t)dM_{i} \equiv \frac{1}{\sqrt{2\pi}}
  \frac{1}{\left[\sigma^{2}(M_{i})-\sigma^{2}(M_{f})\right]^{3/2}}
  \left[-\frac{d\delta_{c}(t)}{dt}\right]
  \left|\frac{d\sigma^{2}(M_{i})}{dM_{i}}\right| dM_{i}.
\end{equation}
As mentioned in $\S$~\ref{secfc}, for a cluster merger with mass greater than
the threshold mass $M_{\mathrm{th}}$ to form a radio halo, the cores of the
progenitor sub-clusters must have been disrupted in the merging process. Assume
a cluster of mass $M$ ($M \geq M_{\mathrm{th}}$) is formed from the merging
process of two sub-clusters $M_{1}$ and $M_{2}$, where $M_{1} \geq M_{2}$. For
$M_{1}$ to have significant core disruption during the merging process, $M_{2}$
have to be greater than some threshold mass, $M_{2} \geq
\triangle_{\mathrm{m}}M_{1}$, according to \citet{sal98}. Since $M_{1} \geq
M_{2}$ and $M=M_{1}+M_{2} \geq (1+\triangle_{\mathrm{m}})M_{1}$, we obtain the
mass range of $M_{1}$: $M/2 \leq M_{1} \leq M/(1+\triangle_{\mathrm{m}})$. The
quantity $M_{2}$ can be the accumulated mass of the merged sub-clusters in
multiple merging. The formation rate of radio halos in cluster mergers with
mass $M \geq M_{\mathrm{th}}$ at time $t$ is given by
\begin{eqnarray}\label{Efr}
  &&R_{f}(M,t) = \int_{M_{a}}^{M_{b}}r(M'\rightarrow M;t)dM',  \nonumber \\
  && = \sqrt{\frac{2}{\pi}} \left[-\frac{d\delta_{c}(t)}{dt}\right]
  \left[\frac{1}{\left[\sigma^{2}(M_{b})-\sigma^{2}(M)\right]^{1/2}} -
  \frac{1}{\left[\sigma^{2}(M_{a})-\sigma^{2}(M)\right]^{1/2}}\right],
\end{eqnarray}
where $M_{a}=M/2$, and $M_{b}=M/(1+\triangle_{\mathrm{m}})$.

\subsection{Cumulative Comoving Number Density}\label{seccnd}

The comoving number density of radio halos that form with cluster mass $M \sim
M+dM$ at time $t_{f} \sim t_{f}+dt_{f}$ and survive until a latter time $t$ is
\begin{eqnarray}\label{Esnd}
  &&n_{\mathrm{rh}}(M,t_{f},t)dMdt_{f} \nonumber \\
  &&=\left\{
  \begin{array}{l@{\quad \mathrm{if} \quad}l}
  n_{\mathrm{ps}}(M,t_{f})R_{f}(M,t_{f})P_{s}(M,t_{f},t)dMdt_{f} &
  t_{f}\leq t \leq t_{f}+t_{\mathrm{rh}}; \\
  0 & t > t_{f}+t_{\mathrm{rh}},
  \end{array}
  \right.
\end{eqnarray}
where $P_{s}$ is the survival probability defined by KS,
\begin{equation}\label{Esp}
  P_{s}(M,t_{1},t_{2}) =
  P(M' < (1+\triangle_{\mathrm{m}})M,t_{2}|M,t_{1}),
\end{equation}
which is the probability that a cluster merger of mass $M$ at $t_{1}$ will
survive to have mass $M'$ less than $(1+\triangle_{\mathrm{m}})M$ at $t_{2}$,
and $t_{\mathrm{rh}}$ is the lifetime of radio halos. The lifetimes of radio
halos may be slightly different from one another, but we ignore the deviation
of the lifetime in calculation for simplicity. A survival merger increases its
mass only by accretion without disrupting its core structure during the
lifetime of its radio halo. We note that a radio halo might survive a new
core-disruption merger even its host cluster are destroyed; the survival radio
halo is treated as a new radio halo possessed by the new merger in our scheme.

The cumulative comoving number density of radio halos with cluster mass $\geq
M(\geq M_{\mathrm{th}})$ at time $t$ can be evaluated as
\begin{eqnarray}\label{EndM}
  &&n_{\mathrm{rh}}(\geq M,t)
  = \int_{M}^{\infty}\int_{t-t_{\mathrm{rh}}}^{t}n_{\mathrm{rh}}(M',t_{f},t)dM'dt_{f} \nonumber \\
  &&+\int_{M_{\mathrm{L}}}^{M}\int_{t-t_{\mathrm{rh}}}^{t} n_{\mathrm{ps}}(M',t_{f})R_{f}(M',t_{f})
  P(M\leq M'' < (1+\triangle_{\mathrm{m}})M',t|M',t_{f})dM'dt_{f},
\end{eqnarray}
where
\[
P(M\leq M'' < (1+\triangle_{\mathrm{m}})M',t|M',t_{f}) = P(M'' \geq M,t|M',t_{f})-P(M'' \geq
(1+\triangle_{\mathrm{m}})M',t|M',t_{f}),
\]
and
\[
  M_{\mathrm{L}} = \left\{
  \begin{array}{l@{\quad \mathrm{if} \quad}l}
  M/(1+\triangle_{\mathrm{m}}) & M > (1+\triangle_{\mathrm{m}})M_{\mathrm{th}}; \\
  M_{\mathrm{th}} & M_{\mathrm{th}} \leq M \leq (1+\triangle_{\mathrm{m}})M_{\mathrm{th}}.
  \end{array}
  \right.
\]
The symbol $M'$ is the cluster mass at $t_{f}$ and $M''$ the cluster mass at
$t$. If $t_{\mathrm{rh}}$ equals to the cosmological time, the term
$t-t_{\mathrm{rh}}$ in the time integral is replaced by 0. The second term in
the right hand side of equation~(\ref{EndM}) represents the number density of
radio halos that form with cluster mass less than $M$ but increase their mass
to greater than $M$ by accretion at the time $t$.

\section{RESULTS AND COMPARISONS WITH OBSERVATIONS}\label{seccom}

\begin{figure}
\plottwo{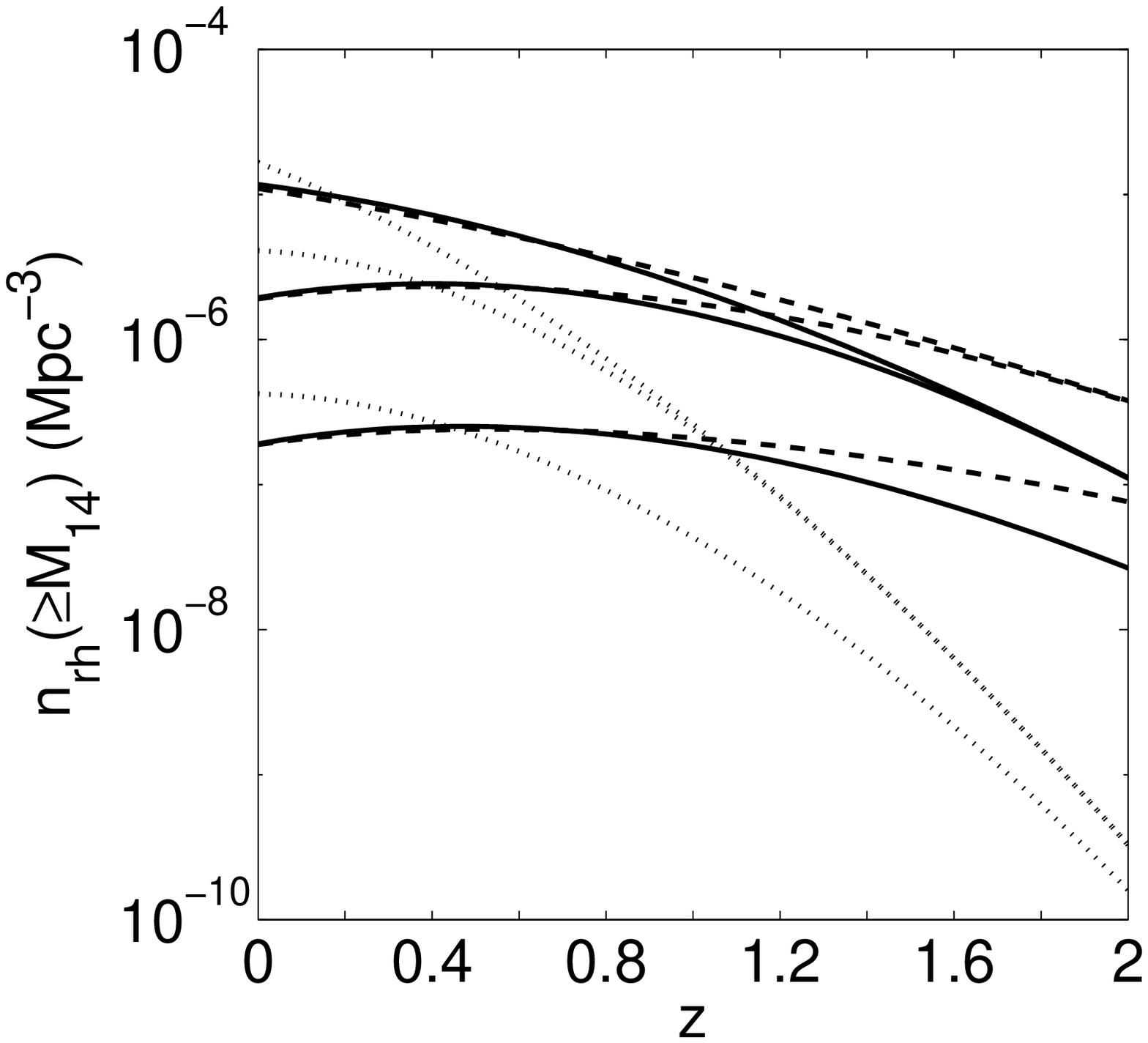}{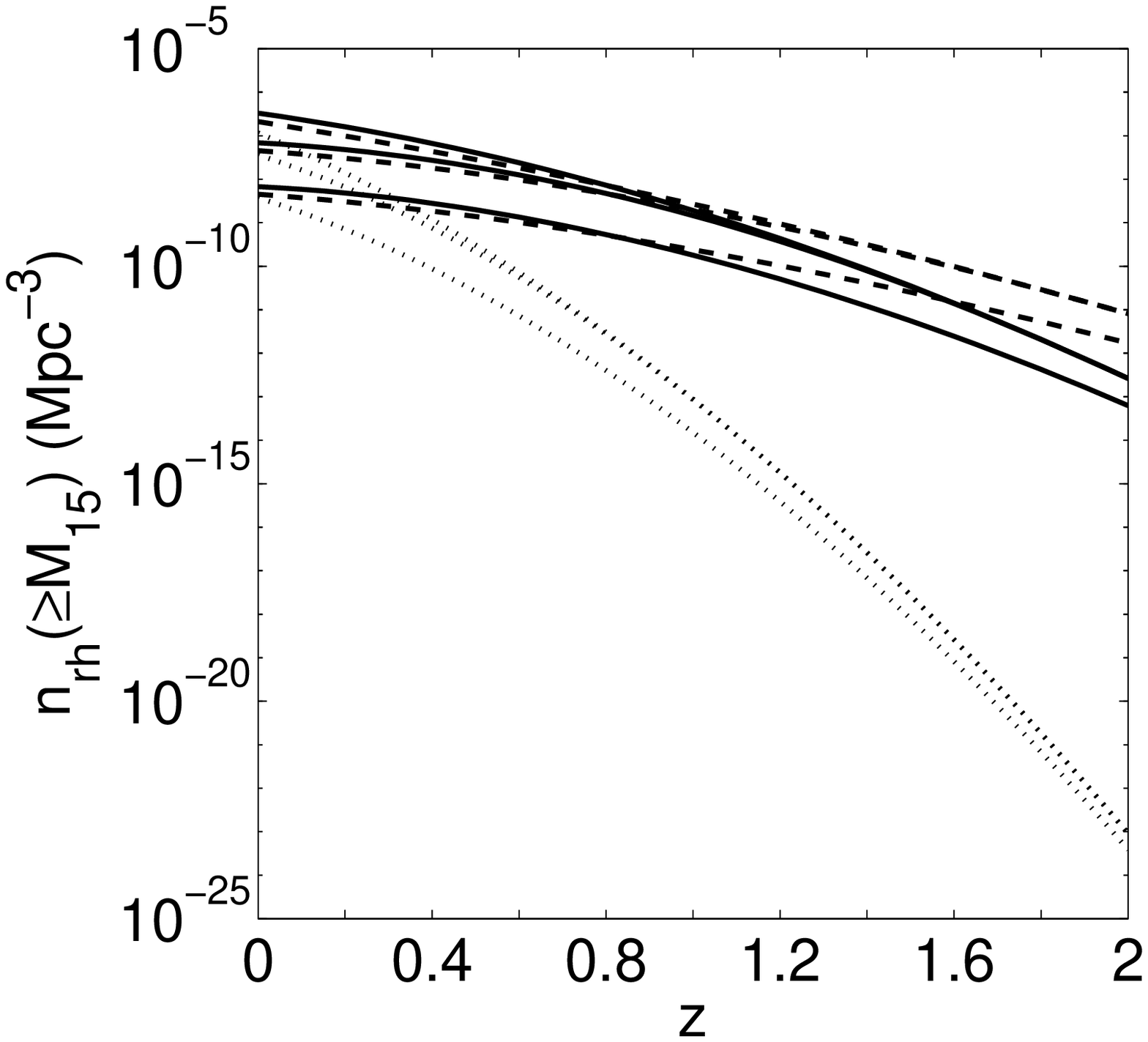}

\caption{(a) The evolution of the total number density of radio halos with cluster mass $M \geq
M_{14}$, where $M_{\mathrm{14}}=10^{\mathrm{14}}M_{\odot}$. Three different cosmological models,
SCDM (\textit{dotted curves}), OCDM (\textit{dashed curves}), and $\Lambda$CDM (\textit{solid
curves}) with three representative lifetimes of radio halos: 0.1 Gyr, 1 Gyr, and the cosmological
time (\textit{bottom to top}) are shown. (b) Same as (a), but for $M \geq
10^{\mathrm{15}}M_{\odot}$.}\label{fig1}

\end{figure}

The evolution of the total comoving number density of radio halos is shown in
Figure~\ref{fig1}. We note that there is a maximum at z $\sim 0.4-0.5$ in the
evolution of the total number density of radio halos for clusters with mass $M
\geq 10^{14}M_{\odot}$ with $t_{\mathrm{rh}}=$ 0.1 Gyr and 1 Gyr in the
$\Lambda$CDM and OCDM models; there is no such distribution maximum if we
consider only the clusters with mass $M \geq 10^{15}M_{\odot}$. This may
represent that before the period, the formation of many clusters with mass
$\geq 10^{14}M_{\odot}$ are significantly due to violent merger of subclusters
with mass $< 10^{14}M_{\odot}$; after the period, the violent-merger rates are
very small and the formation of the clusters with mass $\geq 10^{14}M_{\odot}$
are mainly due to accretion of subclusters with mass $< 10^{14}M_{\odot}$. But
many massive clusters ($M \geq 10^{15}M_{\odot}$) are still mainly due to
violent merger of subclusters.

\begin{figure}
\plotone{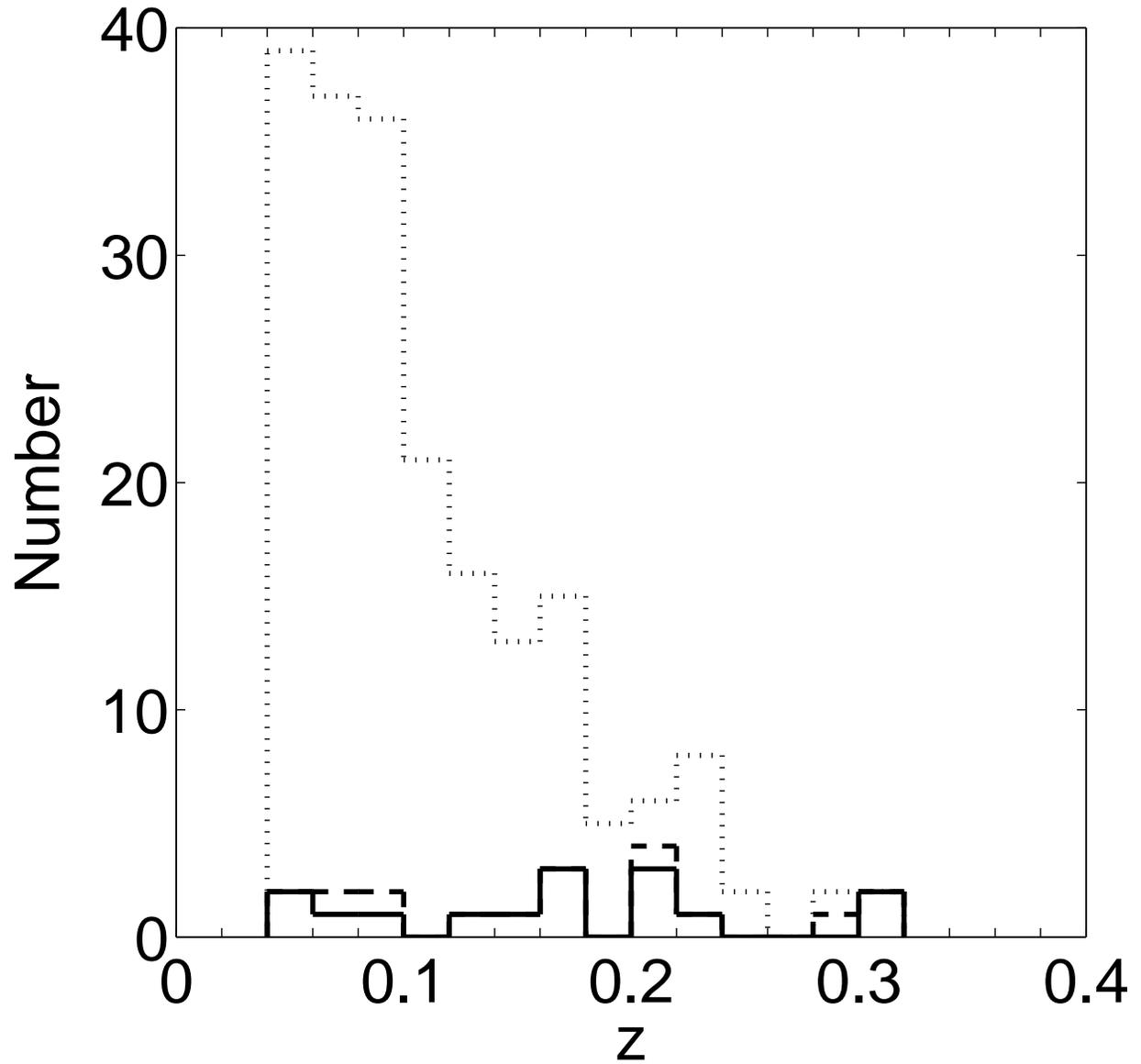}

\caption{The distributions of the X-ray clusters inspected (\textit{dotted line}) and radio halos
detected (\textit{solid line} and \textit{dashed line}) by \citet{gio99}. The \textit{dashed line}
includes halo sources that have uncertainty in their detection.}\label{fig2}

\end{figure}

\citet{gio99} searched radio halo candidates in the NRAO VLA Sky Survey from a
sample of X-ray bright clusters presented by \citet{ebe96}. They found that the
percentage of galaxy clusters possessing diffuse radio sources is $6\% \sim
9\%$ for $L_{\mathrm{X}} \leq 10^{45}$ erg s$^{-1}$ and $27\% \sim 44\%$ for
$L_{\mathrm{X}} > 10^{45}$ erg s$^{-1}$. Here $L_{\mathrm{X}}$ is the
luminosity in the 0.1--2.4 keV energy band. The redshift distributions of radio
halos and the parent X-ray cluster population inspected by \citet{gio99} are
shown in Figure~\ref{fig2}. In this figure, We have corrected two uncertain
halos (A754 and A2219) as confirmed ones and eliminated the uncertain halo in
A2390 which is found to be a mini-halo \citep{bac03}. For $L_{\mathrm{X}} >
10^{45}$ erg s$^{-1}$, the corrected percentage of galaxy clusters possessing
diffuse radio sources is $28\% \sim 41\%$.

In Table~\ref{tbl}, we show the number density ratios of radio halos to
clusters in different cosmological models with different lifetimes of radio
halos. To compare with the observations, we divide the mass of clusters
possessing radio halos into two ranges, $M_{\mathrm{th}} \leq M <
10^{15}M_{\odot}$ and $M \geq 10^{15}M_{\odot}$. The mass $M \sim
10^{15}M_{\odot}$ roughly corresponds to the luminosity $L_{\mathrm{X}} \sim
10^{45}$ erg s$^{-1}$ \citep[e.g.,][]{ros02}. We note that the ratios are a
function of redshift and the percentages in Table 1 are estimated at low z
$\leq 0.4$ so that they can be compared with the observations. Obviously, the
theoretical percentages of radio halos with $t_{\mathrm{rh}}=0.1$ Gyr are much
lower than the observational results for $L_{\mathrm{X}} > 10^{45}$ erg
s$^{-1}$. In other words, we would expect to observe much less radio halos in
high X-ray luminosity clusters if radio halos were transient phenomena with
lifetimes $\sim 0.1$ Gyr. On the other hand, the expected percentages are much
higher than observational results for those radio halos with $t_{\mathrm{rh}}=$
cosmological time; we would have observed much more radio halos if radio halos
had the cosmological lifetime. We find that only radio halos with
$t_{\mathrm{rh}}=1$ Gyr can produce results roughly matching the observations
in two mass ranges. Because of the limits of present instruments, the detection
of radio halos in high-luminosity clusters is easier than that in
low-luminosity ones. Thus the percentage of radio halos for $L_{\mathrm{X}}
> 10^{45}$ erg s$^{-1}$, $28\% \sim 38\%$, may be more robust than that for
$L_{\mathrm{X}} \leq 10^{45}$ erg s$^{-1}$. We note that different cosmological
models have some effects on the calculated ratios but the effects are too small
to cause confusion.

\begin{figure}
\plotone{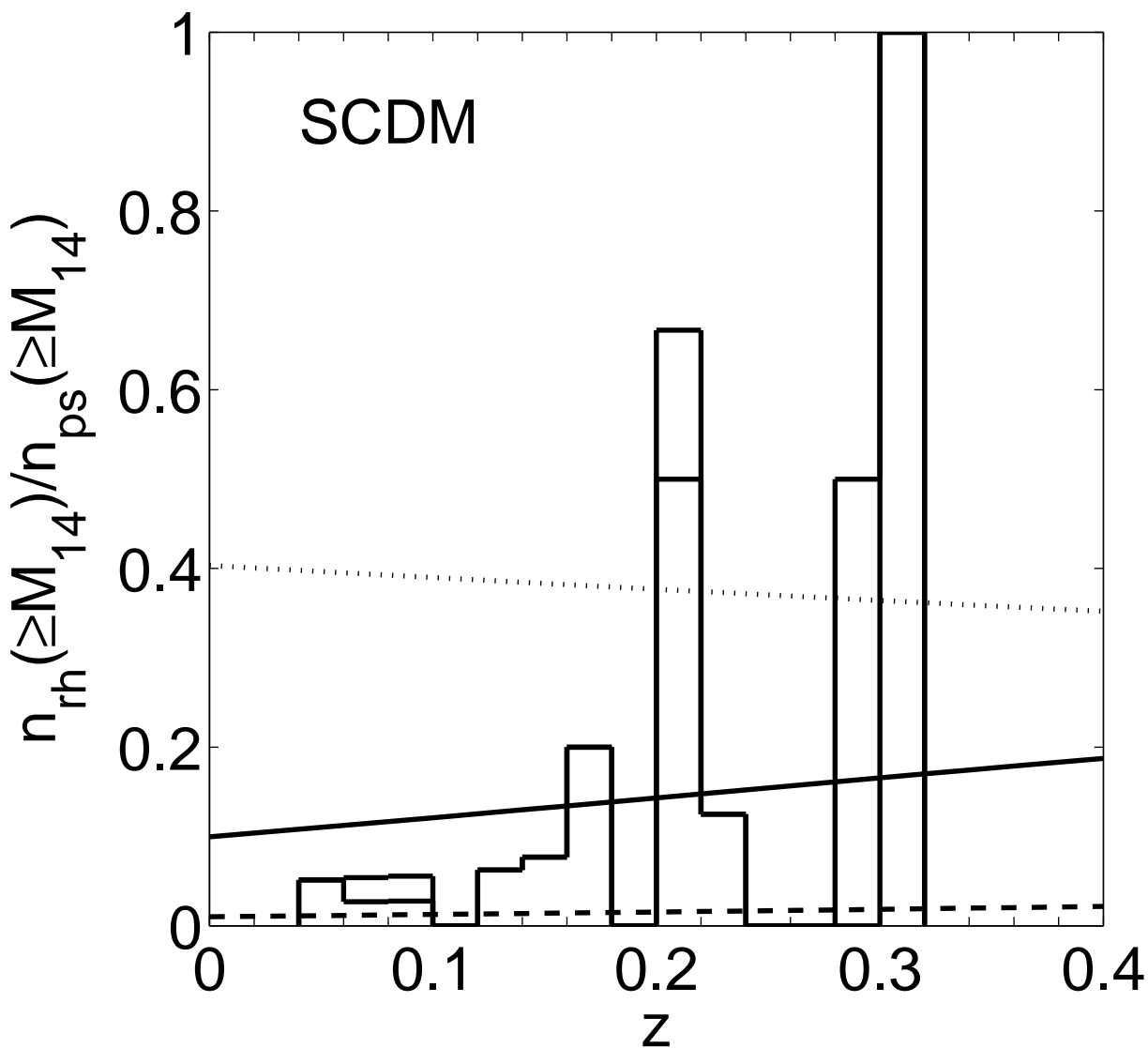}

\caption{The distributions of the ratios of the total number density of radio halos to that of
galaxy clusters in the SCDM model. The results for three representative lifetimes of radio halos:
0.1 Gyr (\textit{dashed curves}), 1 Gyr (\textit{solid curves}), and the cosmological time
(\textit{dotted curves}) are shown. The histograms show the observational results of
Figure~\ref{fig2}; the higher histogram includes the uncertain halos.}\label{fig3}

\end{figure}

\begin{figure}
\plotone{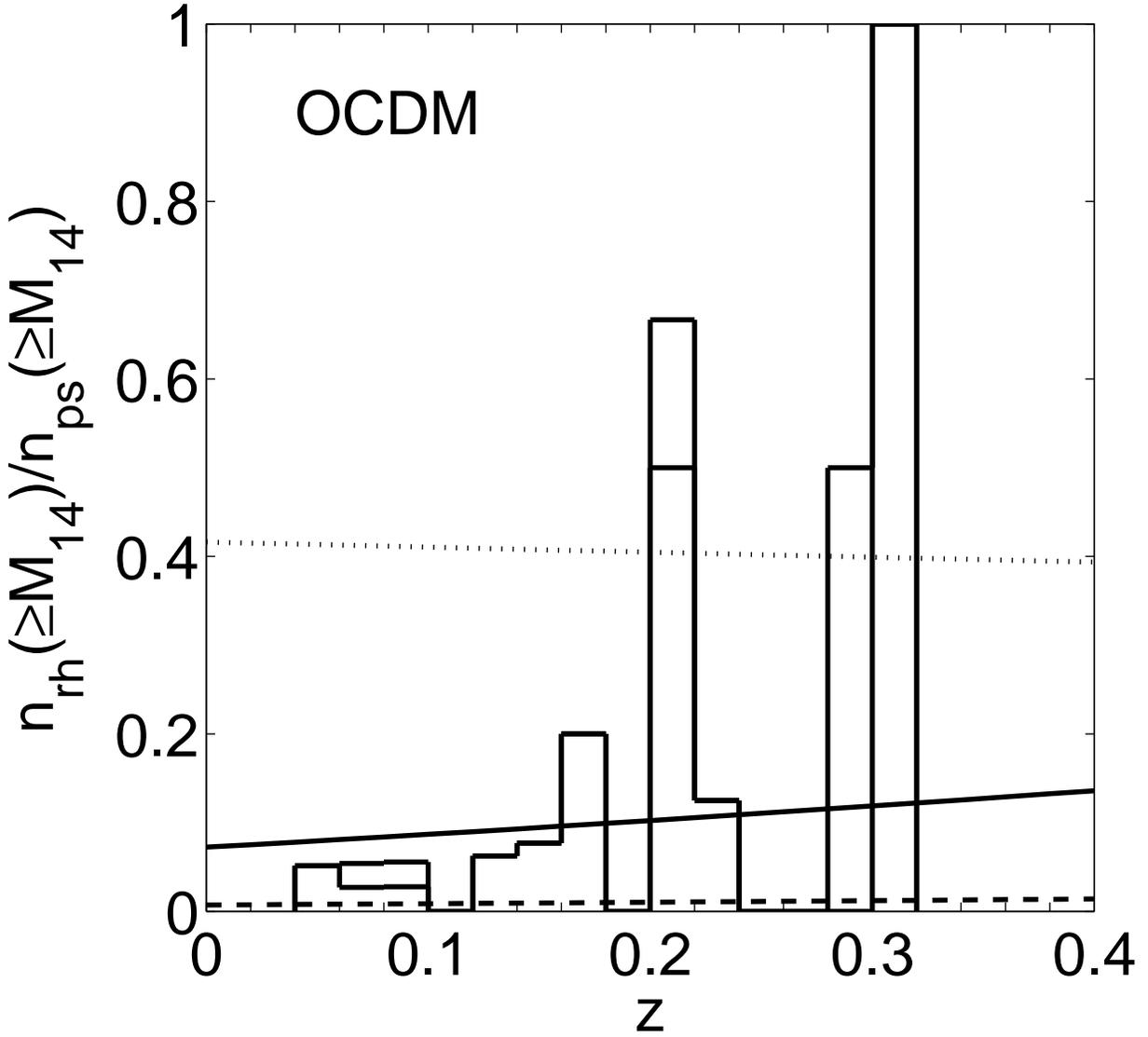}

\caption{Same as Fig.~\ref{fig3}, but for the OCDM model.}\label{fig4}

\end{figure}

\begin{figure}
\plotone{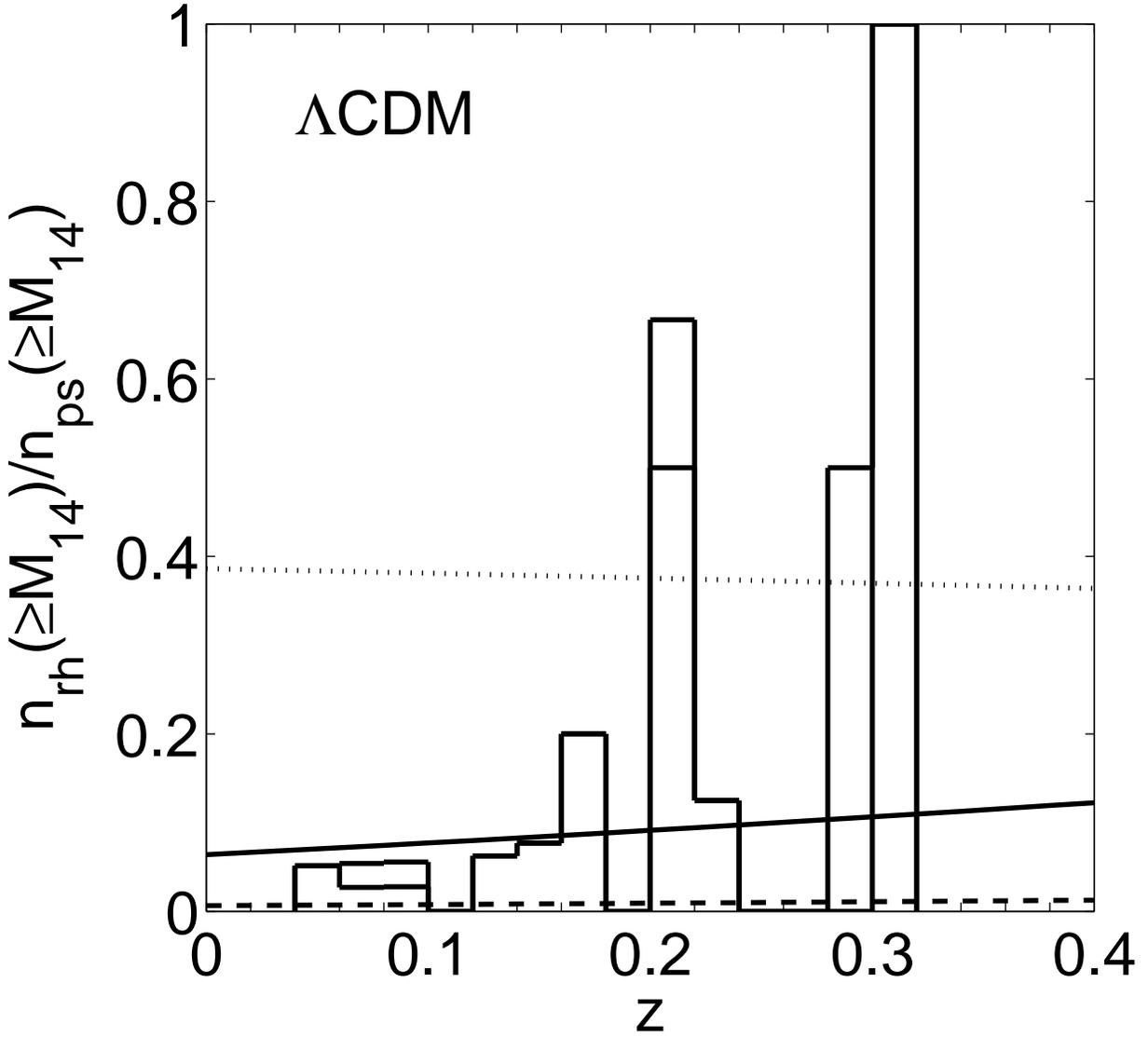}

\caption{Same as Fig.~\ref{fig3}, but for the $\Lambda$CDM model.}\label{fig5}

\end{figure}

In Figures~\ref{fig3}--\ref{fig5}, we show the redshift distributions of the
ratios of radio halos to the galaxy clusters. It is obvious that in all three
different cosmological models only the results with the lifetime
$t_{\mathrm{rh}}=1$ Gyr roughly fit the observations at z $\leq 0.18$. We note
that the number of the cluster sample for z $> 0.18$ are too small and this may
result in a large deviation. In particular, people tend to observe only high
luminous clusters at high redshifts and the ratio of radio halos is higher in
high luminous clusters; the high ratios observed at z $> 0.18$ thus do not
represent the real ratio of radio halos to the total clusters.

\begin{figure}
\plotone{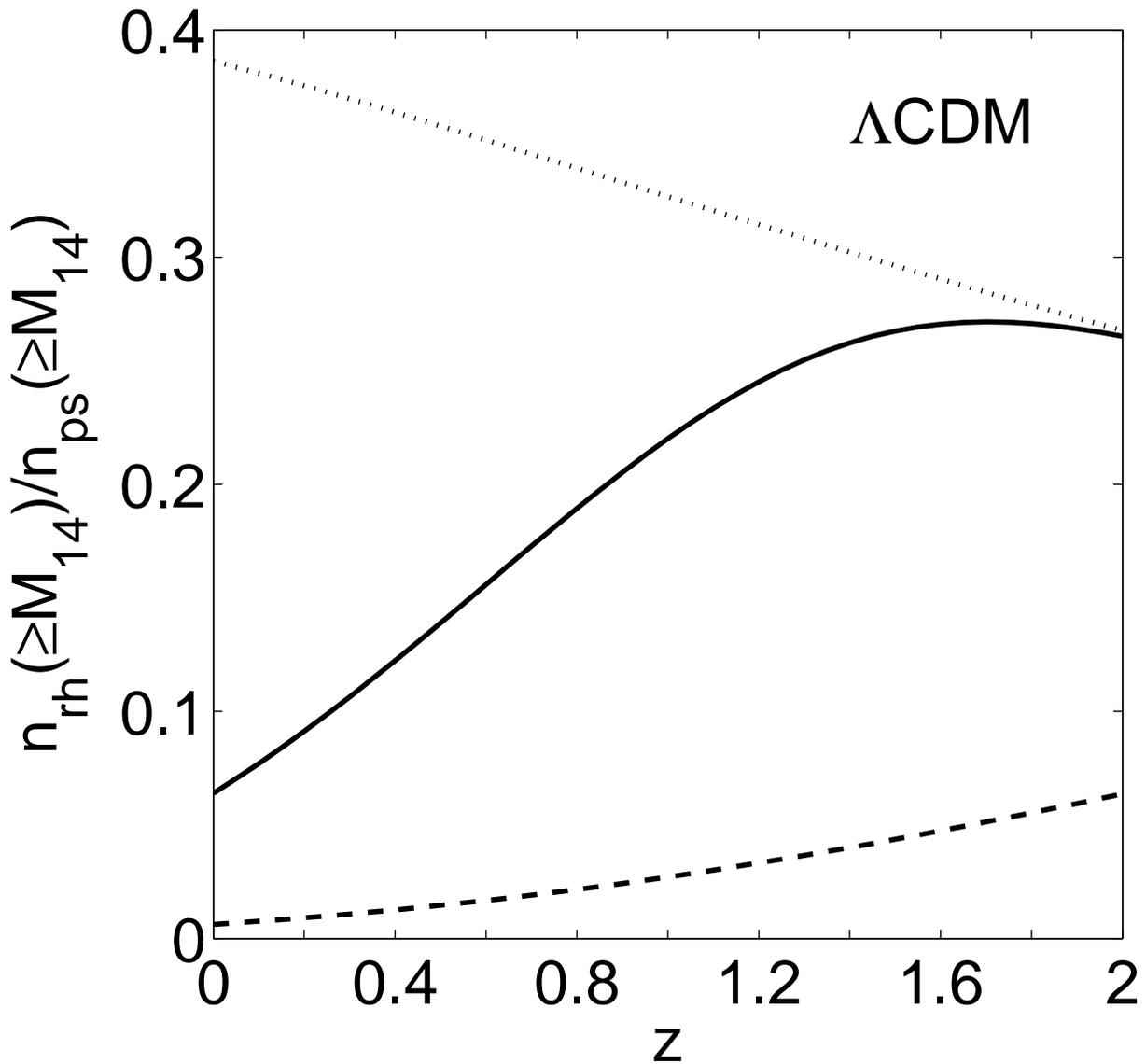}

\caption{The evolution of the ratios of the total number density of radio halos to that of galaxy
clusters in the $\Lambda$CDM model. The results for three representative lifetimes of radio halos:
0.1 Gyr (\textit{dashed curves}), 1 Gyr (\textit{solid curves}), and the cosmological time
(\textit{dotted curves}) are shown.}\label{fig6}

\end{figure}

In Figure~\ref{fig6}, we show the evolution of the ratios up to z $= 2$. For
simplicity, we only show the results of the $\Lambda$CDM models; other
cosmological models show similar results. We find that the ratios of the model
with cosmological lifetime always increase, the model of Gyr lifetime show a
distribution peak at z $= 1.6$, and the ratios of the 0.1 Gyr models decrease
as the universe evolves. These results can also be used as an indicator of the
origin of the radio sources if we have precise measurements and statistics for
clusters at high redshifts.

To investigate the effects of the threshold mass $M_{\mathrm{th}}$, we have
adopted different $M_{\mathrm{th}}$ on our models. In Table~\ref{tb2}, we show
the ratios of radio halos to the galaxy clusters in the mass range
$M_{\mathrm{th}} \leq M < 10^{15}M_{\odot}$ for $M_{\mathrm{th}}=5 \times
10^{13} M_{\odot}$ and $5 \times 10^{14} M_{\odot}$ at z $\leq 0.4$. The ratios
for clusters with mass $\geq 10^{15}M_{\odot}$ are not affected by the low mass
threshold. We find that the ratio distributions of $M_{\mathrm{th}}=5 \times
10^{13} M_{\odot}$ and $5 \times 10^{14} M_{\odot}$ are not very different from
that of $M_{\mathrm{th}}= 10^{14} M_{\odot}$.

\section{DISCUSSION AND CONCLUSIONS}\label{seccon}

The distribution of galaxy clusters possessing radio halos provides a strong
constraint on the origins of radio halos. About $28\% \sim 38\%$ of clusters
with $L_{\mathrm{X}} > 10^{45}$ erg s$^{-1}$ possess radio halos. Since these
X-ray luminous clusters are generally very massive and have undergone through
multiple mergers during their formation, it is expected that these massive
clusters would have accumulated a large amount of cosmic-ray protons from their
formation history \citep[e.g.,][]{min01b}. The cosmic-ray protons would be
confined in the clusters longer than the cosmological time, and the lifetimes
of radio halos would be comparable with the cosmological time in the secondary
electron model. If the secondary electron model was applicable, most of these
massive clusters should possess radio halos; our results show that the
percentage of clusters possessing radio halos should be greater than $70\%$.
This is inconsistent with the observations in which only $\sim$ 35\% of these
massive clusters possess radio halos. According to these results, the secondary
electrons do not seem to be the dominant origin of the radio halos.

On the other hand, if radio halos were transient phenomena associated with a
single acceleration event, such as a major merger shock, they would have
lifetimes $\sim 0.1$ Gyr. Because of the short lifetimes of the sources, radio
halos would be hardly observable even in the massive clusters. The observed
percentage $\sim$ 35\% is thus too high to explain in the hierarchical
clustering formation model.

According to the results presented in $\S$~\ref{seccom}, the lifetimes of radio
halos may be $\sim$ Gyr. As mentioned in $\S$~\ref{secint}, relativistic
electrons in ICM lose energy on the time scale of order $\sim10^{8}$ years
because of the inverse Compton and synchrotron losses. This indicates that a
significant level of re-acceleration is necessary to support the relativistic
electrons against radiative losses and to maintain radio halos to last for
$\sim$ Gyr \citep{bru01,kuo03}.

As discussed in \S~\ref{seccom}, the percentage of radio halos for
$L_{\mathrm{X}} > 10^{45}$ erg s$^{-1}$, $\sim$ 28\%--38\%, is
more robust. We here investigate the effects of the lifetime of
the radio halos, the dividing mass, and the threshold mass ratio
$\triangle_m$ on our results. Only the $\Lambda$CDM model is
considered. First, the lifetimes of radio halos can affect the
ratios of the radio halos. The percentage of radio halos with
$t_{\mathrm{rh}}=1$ Gyr is $\sim 21\%$ and seems to be lower than
the observational results, $\sim$ 28\%--38\%, as shown in
Table~\ref{tbl}. However, we note that the lifetime of the radio
halo is of $\sim$ Gyr and could be slightly longer or shorter than
1 Gyr and the difference of the assumed lifetime could affect the
predicted percentage of the radio halo. For example, if a longer
lifetime for the radio halo, $t_{\mathrm{rh}}=1.5$ Gyr, is
assumed, then the percentage of the radio halo would become $\sim
32\%$ and would be in agreement with the observational results.
Second, the dividing mass at $M = 10^{15} M_{\odot}$ may be lower
than the realistic mass corresponding to $L_{\mathrm{X}} =
10^{45}$ erg s$^{-1}$. For example, the A773 cluster with
$L_{\mathrm{X}} \sim 1.01 \times 10^{45}$ erg s$^{-1}$ has mass in
the range 1.25--2.08 $\times 10^{15} M_{\odot}$ \citep{gov01}. In
Table~\ref{tb3}, we show the percentages corresponding to
different dividing masses. Obviously, the results with
$t_{\mathrm{rh}}=1$ Gyr are improved but those with
$t_{\mathrm{rh}}=$ cosmological time are worse if a higher and
more realistic dividing mass is taken. Third, different threshold
$\triangle_{\mathrm{m}}$ also affect our results. The results with
different $\triangle_{\mathrm{m}}$ are shown in Table~\ref{tb4}.
The values of $\triangle_{\mathrm{m}}$ strongly affect the results
for $t_{\mathrm{rh}}=1$ Gyr and $t_{\mathrm{rh}}=$ cosmological
time. For $\triangle_{\mathrm{m}}=0.7$, the results with
$t_{\mathrm{rh}}=$ cosmological time seems to be close to the
observational results; this implies that the radio halos would
only be generated in the mergers with two nearly equal-mass
progenitors if the secondary electrons were the dominant origin
for forming radio halos. For $\triangle_{\mathrm{m}}=0.5$, the
ratio of the radio halos with $t_{\mathrm{rh}}=1$ Gyr is
consistent with the observational results. Obviously, the
$\triangle_{\mathrm{m}}$ is a dominant factor in the determining
the ratio of the radio halos. Thus it is important to investigate
in detail the exact value of $\triangle_{\mathrm{m}}$ to clarify
the question.

Note that we have assumed that a merger with mass greater than a
threshold mass will form a radio halo if its relative mass
increase exceeds a threshold, $\triangle_{\mathrm{m}}$. It might
be possible that some mergers that satisfy our criteria did not
form radio halos. However, according to the study of \citet{buo01}
for $\sim$ 30 bright X-ray clusters, the number of the unaccounted
mergers should be very small and thus has little effect on our
results.

Magnetic fields could be also an important parameter in
determining the radio powers of the cluster radio halos. It is
very difficult and complicated to determine this parameter.
Magnetic fields $\leq$ 0.4 $\mu$G were found for several clusters
using the inverse Compton models for the hard X-ray excess
\citep{fus03}. \citet{cla01} found that the cluster fields are
typically around 5--10 $\mu$G using rotation measurements.
Furthermore, during a cluster merging process, the magnetic fields
might be amplified by a factor of $\sim$ 20 on small scales
\citep*{roe99} and might have a non-negligible effect on the radio
emission. However, we note that both the magnetic field energy and
the relativistic particle energy are provided by the merging
energy, which is a function of the merger mass. In other words,
the cluster mass should be a more fundamental parameter in
determining the radio powers of the radio halos
\citep{buo01,gov01}. Since we have considered the merger mass as a
main parameter in our calculation, we thus do not treat the
magnetic field as an independent parameter.

We could try to match the results obtained from the secondary
electron model ($\sim$ 70\%) with the observations ($\sim$ 35\%)
by assuming that about half of the radio halos produced by the
secondary electrons are unobservable. However, \citet{min01a}
showed that for a cluster of given mass the radio power from the
secondary electrons could vary by almost an order of magnitude in
their simulation. We note that such level of variation, as also
detected in observation \citep{bac03}, is not large enough to make
a detectable radio halo source to become non-detectable,
especially for high X-ray luminosity clusters at low redshifts. It
is thus unlikely to apply some observational effects to lower the
radio halo fraction of the secondary electron model to match the
observational results.

The radio halo fraction of galaxy clusters have been estimated in some recent
studies. \citet{fuj01} estimated the fraction of cluster radio halos based on
the radiative energy loss timescale of the relativistic electrons. This
timescale is similar to our case for primary electrons without re-acceleration
and can account for only $\sim 10\%$ of observations. To match the
observations, they assumed that even a rather weak merger should also trigger a
radio halo. This assumption seems to be in contradiction with observations,
which showed that only massive clusters experiencing violent mergers can have
halos \citep{buo01} as we have discussed in $\S$~\ref{secfc}. \citet{ens02}
estimated the cluster radio halo luminosity function by assuming a radio halo
fraction $f_{\mathrm{rh}} = 1/3$ for all clusters. However, observations showed
that the halo fraction is very different for low and high luminosity X-ray
clusters and it is inconsistent with current observations to assume a constant
radio halo fraction for all clusters. \citet{gio99} have noted that the lack of
diffuse radio sources in low X-ray luminosity clusters is real because of their
low redshifts; \citet{bac03} have also stressed that the correlation between
the halo radio power and the cluster X-ray luminosity is only applied to
clusters showing major mergers and therefore cannot be generalized to all
clusters. The other more realistic fraction adopted by \citet{ens02} is
estimated by assuming that a cluster can possess a radio halo if its mass
increases by more than 40\% of its present mass within half a dynamical
timescale. In their work, the fraction for clusters with mass of
$10^{15}M_{\odot}$ is 0.32. This seems to agree with the observational results
for $L_{\mathrm{X}} > 10^{45}$ erg s$^{-1}$. However, the fraction for clusters
with mass of $10^{14}M_{\odot}$ and that for clusters with mass of
$10^{13}M_{\odot}$ are 0.26 and 0.22, respectively. These results indicate that
the radio halo ratio for $L_{\mathrm{X}} \leq 10^{45}$ erg s$^{-1}$ should be
greater than 20\% and therefore are in conflict with the observations. We also
note that the effects of the secondary electron model on the radio halo
fraction were not considered in previous studies.

Our results are based on the assumption that the diffuse radio emission is
associated with cluster merger events. We note that radio galaxies, starbursts,
and AGNs might also contribute significant relativistic electrons to the ICM of
the clusters. However, the high-energy ($\geq 1$ GeV) relativistic electrons
injected by these sources would become invisible in a very short time and can
not directly account for the extended radio emission. On the other hand, the
low-energy ($\leq 1$ GeV) relativistic electrons can survive a longer time and
become more extended via diffusion. A cluster merger event, as suggested in
this paper for the diffuse radio emission phenomena, can produce merger shocks
and MHD turbulence that can re-accelerate the survived low-energy relativistic
electrons and reignite the radio emission in a more extended region. In this
respect, our results are also applicable even the relativistic electrons are
originally injected from radio galaxies, starburst galaxies, or AGNs.

We note that the effects of the inverse Compton loss rising with increasing
redshift may reduce the ratio of radio halos at high redshifts, but the results
at low redshifts are only slightly affected and the conclusions are still
tenable.

\acknowledgments

We thank an anonymous referee for helpful comments. This work was
partially supported by the National Science Council of Taiwan
under NSC 92-2112-M-008-023, and by the Ministry of Education of
Taiwan through the CosPA Project 92-N-FA01-1-4-5.

\newpage

\begin{deluxetable}{lccccccc}
\tablecaption{The Ratios of Galaxy Clusters Containing Diffuse
Radio Sources\label{tbl}} \tablehead{Observations &
\multicolumn{3}{c}{$L_{\mathrm{X}} \leq 10^{45}$ erg s$^{-1}$} & &
\multicolumn{3}{c}{$L_{\mathrm{X}} > 10^{45}$ erg s$^{-1}$}\\
\hline\\
 Halos & \multicolumn{3}{c}{$\sim 4\%$} & & \multicolumn{3}{c}{$\sim 28$--$38\%$}\\
 Halos+Relics & \multicolumn{3}{c}{$\sim 6$--$9\%$} & & \multicolumn{3}{c}{$\sim 28$--$41\%$}\\
\hline\\
Models & \multicolumn{3}{c}{$M_{\mathrm{th}}\leq M <
10^{15}M_{\odot}$} & &
\multicolumn{3}{c}{$M \geq 10^{15}M_{\odot}$}\\
\hline\\
 & SCDM & OCDM & $\Lambda$CDM & & SCDM & OCDM & $\Lambda$CDM}
\startdata
0.1 Gyr & $\sim 2\%$ & $\sim 1\%$ & $\sim 1\%$ & & $\sim 4\%$ & $\sim 2\%$ & $\sim 2\%$\\
1 Gyr & $\sim 14\%$ & $\sim 10\%$ & $\sim 9\%$ & & $\sim 38\%$ & $\sim 24\%$ & $\sim 21\%$\\
cosmological time & $\sim 38\%$ & $\sim 40\%$ & $\sim 37\%$ & &
$\sim 80\%$ & $\sim 77\%$ & $\sim
70\%$\\
\enddata

\end{deluxetable}

\clearpage
\begin{deluxetable}{lccccccc}
\tablecaption{The Ratios of Galaxy Clusters Containing Diffuse Radio Sources in
the Mass Range $M_{\mathrm{th}}\leq M < 10^{15}M_{\odot}$ Under Different
$M_{\mathrm{th}}$\label{tb2}} \tablehead{$M_{\mathrm{th}}$ &
\multicolumn{3}{c}{$5 \times10^{14}M_{\odot}$} & &
\multicolumn{3}{c}{$5 \times 10^{13}M_{\odot}$}\\
\hline\\
 & SCDM & OCDM & $\Lambda$CDM & & SCDM & OCDM & $\Lambda$CDM}
\startdata
0.1 Gyr & $\sim 3\%$ & $\sim 2\%$ & $\sim 2\%$ & & $\sim 1\%$ & $\sim 1\%$ & $\sim 1\%$\\
1 Gyr & $\sim 22\%$ & $\sim 16\%$ & $\sim 14\%$ & & $\sim 12\%$ & $\sim 8\%$ & $\sim 8\%$\\
cosmological time & $\sim 39\%$ & $\sim 41\%$ & $\sim 38\%$ & & $\sim 35\%$ & $\sim 37\%$ & $\sim
34\%$\\
\enddata

\end{deluxetable}

\clearpage
\begin{deluxetable}{lccccccccccccccc}
\tablecaption{The Ratios of Galaxy Clusters Containing Diffuse Radio Sources in
the Mass Range $M \geq M'$ in $\Lambda$CDM\label{tb3}} \tablehead{$M'$ &
\multicolumn{3}{c}{$1 \times 10^{15}M_{\odot}$} & & \multicolumn{3}{c}{$1.2
\times 10^{15}M_{\odot}$} & & \multicolumn{3}{c}{$1.5 \times 10^{15}M_{\odot}$}
& & \multicolumn{3}{c}{$2 \times 10^{15}M_{\odot}$}
\\} \startdata
0.1 Gyr & \multicolumn{3}{c}{$\sim 2\%$} & & \multicolumn{3}{c}{$\sim 3\%$} & & \multicolumn{3}{c}{$\sim 3\%$} & & \multicolumn{3}{c}{$\sim 3\%$} \\
1 Gyr & \multicolumn{3}{c}{$\sim 21\%$} & & \multicolumn{3}{c}{$\sim 23\%$} & & \multicolumn{3}{c}{$\sim 25\%$} & & \multicolumn{3}{c}{$\sim 29\%$} \\
cosmological time & \multicolumn{3}{c}{$\sim 70\%$} & & \multicolumn{3}{c}{$\sim 73\%$} & & \multicolumn{3}{c}{$\sim 77\%$} & & \multicolumn{3}{c}{$\sim 82\%$} \\
\enddata

\end{deluxetable}

\begin{deluxetable}{lccccccccccc}
\tablecaption{The Ratios of Galaxy Clusters Containing Diffuse Radio Sources
for Different $\bigtriangleup_{\mathrm{m}}$ in the Mass Range $M \geq
10^{15}M_{\odot}$ in $\Lambda$CDM\label{tb4}}
\tablehead{$\bigtriangleup_{\mathrm{m}}$ & \multicolumn{3}{c}{0.5} & &
\multicolumn{3}{c}{0.6} & & \multicolumn{3}{c}{0.7}
\\} \startdata
0.1 Gyr & \multicolumn{3}{c}{$\sim 3\%$} & & \multicolumn{3}{c}{$\sim 2\%$} & & \multicolumn{3}{c}{$\sim 2\%$} \\
1 Gyr & \multicolumn{3}{c}{$\sim 32\%$} & & \multicolumn{3}{c}{$\sim 21\%$} & & \multicolumn{3}{c}{$\sim 14\%$} \\
cosmological time & \multicolumn{3}{c}{$\sim 95\%$} & & \multicolumn{3}{c}{$\sim 70\%$} & & \multicolumn{3}{c}{$\sim 49\%$} \\
\enddata

\end{deluxetable}

\clearpage
\newpage

\clearpage

\clearpage

\clearpage

\clearpage

\clearpage

\end{document}